\documentclass[final,3p,times,11pt]{elsarticle}

\usepackage{amssymb}

\usepackage{amsmath}
\usepackage{booktabs} 
\usepackage{bm}
\usepackage[export]{adjustbox}
\usepackage{multirow}
\usepackage{graphicx}
\usepackage{caption}
\usepackage{subcaption}
\journal{Mechanical Systems and Applications}

\usepackage{xcolor}

\begin{document}

\begin{frontmatter}



\title{Benchmarking deep learning models for bearing fault diagnosis using the CWRU dataset: A multi-label approach}


\author[inst1]{Rodrigo Kobashikawa Rosa}

\affiliation[inst1]{organization={Department of Electrical and Electronic Engineering, Federal University of Santa Catarina},
            addressline={Campus Universitário Reitor João David Ferreira Lima, s/nº}, 
            city={Florianópolis},
            postcode={88040-900}, 
            state={Santa Catarina},
            country={Brazil}}
            
\author[inst2]{Danilo Braga}

\author[inst1]{Danilo Silva}

\affiliation[inst2]{organization={Research and Innovation Department, Dynamox},
            addressline={Rua Coronel Luiz Caldeira, nº 67, bloco C - Condomínio Ybirá}, 
            city={Florianópolis},
            postcode={88034-110}, 
            state={Santa Catarina},
            country={Brazil}}

\begin{abstract}
This paper proposes a novel approach for modeling the problem of fault diagnosis using the Case Western Reserve University (CWRU) bearing fault dataset. Although the dataset is considered a standard reference for testing new algorithms, the typical dataset division suffers from data leakage, as shown by Hendriks et al. (2022) and Abburi et al. (2023), leading to papers reporting over-optimistic results. While their proposed division significantly mitigates this issue, it does not eliminate it entirely. Moreover, their proposed multi-class classification task can still lead to an unrealistic scenario by excluding the possibility of more than one fault type occurring at the same or different locations. As advocated in this paper, a multi-label formulation (detecting the presence of each type of fault for each location) can solve both issues, leading to a scenario closer to reality. Additionally, this approach mitigates the heavy class imbalance of the CWRU dataset, where faulty cases appear much more frequently than healthy cases, even though the opposite is more likely to occur in practice. A multi-label formulation also enables a more precise evaluation using prevalence-independent evaluation metrics for binary classification, such as the ROC curve. Finally, this paper proposes a more realistic dataset division that allows for more diversity in the training dataset while keeping the division free from data leakage. The results show that this new division can significantly improve performance while enabling a fine-grained error analysis. As an application of our approach, a comparative benchmark is performed using several state-of-the-art deep learning models applied to 1D and 2D signal representations in time and/or frequency domains.
\end{abstract}




\end{frontmatter}


\section{Introduction}
\label{sec:introduction}
Detecting and diagnosing mechanical failures in industrial machines is crucial for maintaining a continuous production process. Within the domain of rotating machinery, rolling element bearings are numerous, representing most of the bearings, while being critical components and often the primary source of mechanical failures \cite{zhang2020deep}. Therefore, adopting automated techniques is strategic for early fault detection, enabling maintenance scheduling and reducing downtime and operational costs. In this sense, vibration sensors offer a promising data-driven approach to detecting bearing faults, significantly aiding in the precise and timely identification of potential issues in rotating machinery. Nevertheless, with a large number of machines monitored by vibration sensors, manual analysis by specialists is not scalable. Alternatively, quick analysis of large data volumes is achievable through automated fault analysis techniques, ensuring potential problems are detected and addressed promptly. Among these techniques, machine learning methods have been successfully applied \cite{lei2020applications, zhang2019deep}, with deep learning algorithms \cite{neupane2020bearing} achieving state-of-the-art performance. However, employing and benchmarking these techniques requires a careful methodology \cite{kapoor2023leakage}, problem formulation, and dataset preparation to ensure they generalize to real-world scenarios.

The CWRU bearing fault dataset \cite{cwru} has become a widely used benchmark dataset for new network architectures \cite{neupane2020bearing}. It comprises vibration signals collected during experiments on a test bench featuring rolling bearings within an electric motor. The dataset encompasses bearings with single-point faults of varying sizes in the inner race, outer race, and rolling element (ball), as well as healthy bearings. Measurements were obtained from both the drive end and fan end locations, with healthy and faulty bearings installed under four operational motor load conditions for each fault type. As a result, the experiments feature several bearing configurations for faulty bearings and just one for healthy bearings. This disproportion leads to a heavy skew towards abnormality in the dataset, which is a somewhat unrealistic reflection of the more prevalent healthy condition observed in real-world scenarios.

While abundant in the literature, it is unclear if fault classification results obtained with the CWRU dataset using machine learning models truly generalize to different settings. The typical CWRU dataset division seen in the literature has the segments from the same signal being randomly split between train and test sets. 
This creates a situation known as \textit{data leakage} \cite{kapoor2023leakage} that can cause the machine learning model to memorize a signature of each specific signal, compromising the generalization of the model to new unseen signals. It is especially troublesome for deep learning models due to their capacity to easily overfit, resulting in an over-optimistic evaluation that would fail to be reproduced in practice.
Another typical CWRU division, mostly used in the context of domain adaptation, is the division of the dataset by motor load. In this division, the aforementioned segment-level data leakage is no longer present 
because all segments from the same signal remain together in either the train or the test set.
However, both Hendriks et al. \cite{hendriks2022towards} and Abburi et al. \cite{abburi2023closer} identified another form of data leakage present in this division.
In the CWRU dataset, the same faulty bearings are reused for experiments at different loads, so that multiple measurements are made for each bearing. As a result, if the train and test datasets are divided by load, measurements at different loads are created from the same faulty bearings, resulting in the same bearing appearing in both the train and test datasets. 
This situation is also a form of data leakage, as the bearing information is leaked between train and test sets. Indeed, both papers show that it leads to over-optimistic results when comparing to divisions where this leakage does not occur. 

Although their methodology has effectively solved most of the data leakage problem and presented a more practical scenario than previous studies, some disadvantages remain. The multi-class formulation present in both works comes with several limitations, the most prominent being its inability to operate completely without data leakage due to the presence of only one healthy bearing configuration in the dataset, making it impossible to split the dataset and prevent measurements from the same healthy bearing from appearing in both train and test datasets. Additionally, since classes are defined based on the CWRU bearing configurations, where at most a single fault is present on each configuration, the model cannot account for multiple faults happening at once.
Moreover, the highly imbalanced healthy class renders accuracy an unrealistic evaluation metric. Finally, in the Hendriks et al. \cite{hendriks2022towards} paper, the authors rely on the use of two synchronous accelerometers to identify the fault location, which is an unlikely assumption for real-case scenarios and unnecessary for fault detection and diagnosis.

This paper proposes a novel approach for modeling the problem using binary multi-label classification. 
Specifically, for each location (fan end and drive end), a model must detect the presence or absence of each type of fault (inner, outer and ball).
With this approach, it is possible to completely eliminate the data leakage issue by ensuring that all signals from the healthy bearing pair appear only in the test set while also leading to a more realistic scenario by having the possibility of more than one fault type occurring at each motor bearing. The multi-label approach also mitigates the healthy bearing class imbalance by transforming the problem into a separate binary classification for each of the three fault types, making the negative class of each fault-type detector comprise the other fault types instead of only the healthy class. Consequently, the approach enables a more precise evaluation by using prevalence-independent metrics such as the ROC curve and the AUROC. Moreover, by transforming the problem into two separate problems (one at each possible fault location) and, for each problem, evaluating only a single signal (the signal acquired at that specific location), it is possible to eliminate the need for synchronous signals.
Lastly, with this multi-label approach, a different dataset division is made possible that maximizes the diversity of fault types and sizes during training and testing while keeping it without data leakage. 

Experiments under our proposed methodology were conducted on the ResNet18 architecture applied to raw spectrogram images. A rigorous performance evaluation is presented, along with an error analysis of the model's outputs, as well as ablation experiments to evaluate the gains of individual components of our proposed training approach and dataset division. Additionally, we conducted a comparative benchmark using several state-of-the-art deep learning models applied to different signal representations to identify the most suitable architecture. Finally, we also consider a simpler problem of detecting when a fault (of any type) is present at a specific location, for which performance can be further improved.

\section{Related work}

To address the bearing information from leaking between splits, Hendriks et al. \cite{hendriks2022towards} proposed a new dataset split. Using the same idea of the typical division where the sets are split by load, the authors noted that if the split was made by fault size, the bearing information leakage no longer occurred for the faulty bearing measurements. Considering the problem of fault classification into seven classes (healthy state plus inner/outer/ball faults at either fan or drive end), their proposed approach significantly reduced the inflated performance metric of state-of-the-art deep convolutional networks, which achieved 95\% classification accuracy in the division by load and now achieve a more realistic performance of 53\% accuracy in the division by fault size.

Abburi et al. \cite{abburi2023closer}, on the other hand, proposed a different dataset split to mitigate bearing information from leaking. In their proposal division, they assign only drive-end fault measurements for the training set, fan-end fault measurements of sizes 7 and 14 mils for the validation set, and fan-end measurements of size 21 mils for the test sets. The experiments used traditional machine learning models, such as Random Forest, Naive Bayes, and Support Vector Machine, on a multi-class problem formulation with three fault classes (inner race, outer race, ball) and the healthy condition. The results have shown that the bearing split had worse results across all metrics reported (accuracy, precision, macro F1-score, recall) when compared to the random split where the bearing information leakage occurred. For the Naive Bayes model, the accuracy dropped from 85.8\% to 69.5\%.

However, as mentioned earlier, none of these multi-class formulations is able to avoid bearing data leakage from the healthy class, which we avoid using a multi-label approach.

Although a few previous works using the CWRU dataset for fault detection and diagnosis have also adopted a multi-label approach, they all employ different formulations that ignore the specific problems addressed in this paper.
Shen et al. \cite{shen2020deep} consider two multi-class labels representing fault type and fault size in the CWRU dataset. Since a single label is used for fault type diagnosis, this is still a multi-class problem, where fault types are assumed to be mutually exclusive.
Similarly, Yu et al. \cite{yu2021multi} consider three multi-class labels corresponding to fault size, fault type and motor speed, except that the healthy condition is ignored. Again this is a multi-class problem for fault type diagnosis, while the fault detection problem is not treated.
The closest to our formulation are those of Chen et al. \cite{chen2019multi} and Jin et al. \cite{jin2021actual}, which consider binary labels for the three fault types plus the healthy state (as well as labels for each fault type and size combination in \cite{chen2019multi}). However, the inclusion of a healthy state label requires using healthy data for training (besides testing), which our approach does not require. Moreover, these works only consider signals acquired at a single location (either drive end or fan end) corresponding to faults occurring at that same location; thus, the fault localization problem is not treated. 

More importantly, these works differ from ours in their division of the dataset into train and test splits, which fails to avoid data leakage \cite{abburi2023closer}, either at the segmentation level or at the bearing level. 
These leakages predominantly cause overoptimistic results with nearly 100\% accuracy, and therefore, their performance cannot be properly evaluated. 
Another point not present in any of these studies is the use of the fan end signals as negative samples when diagnosing faults at the drive end and vice-versa. 
To the best of our knowledge, our paper is the only study that has applied a multi-label approach to the CWRU dataset to address problems such as bearing data leakage and healthy class imbalance, ultimately transforming the problem formulation into one that is better aligned with real-world conditions while ensuring proper evaluation of the model.

\section{Methodology}
\label{sec:methodology}

\subsection{Dataset}
\label{sec:dataset}
The CWRU bearing fault dataset \cite{cwru} is a collection of experiments that involved a single pair of healthy bearings and several artificially created faulty bearings. The faults were created through electro-discharge machining, introducing point faults with diameters of 7, 14, 21, and 28 mils in the inner race, outer race, and rolling element separately. For the outer race faults, the experiments considered faults located at three different positions relative to the load zone. The healthy and faulty bearings were reinstalled at both the drive end (DE) and fan end (FE) locations (where each configuration comprises either two healthy bearings or one healthy bearing and one faulty bearing), and data were collected synchronously, with one accelerometer at each location. For each configuration, experiments were made using four operational motor load conditions ranging from 0 (no load) to 3 horsepower (HP). In most cases, the experiments used a 12 kHz sampling rate, while some used 48 kHz. All the experiments consist of signals that are approximately 10 seconds long.

\begin{figure*}[ht]
    \centering
    \includegraphics[width=\linewidth]{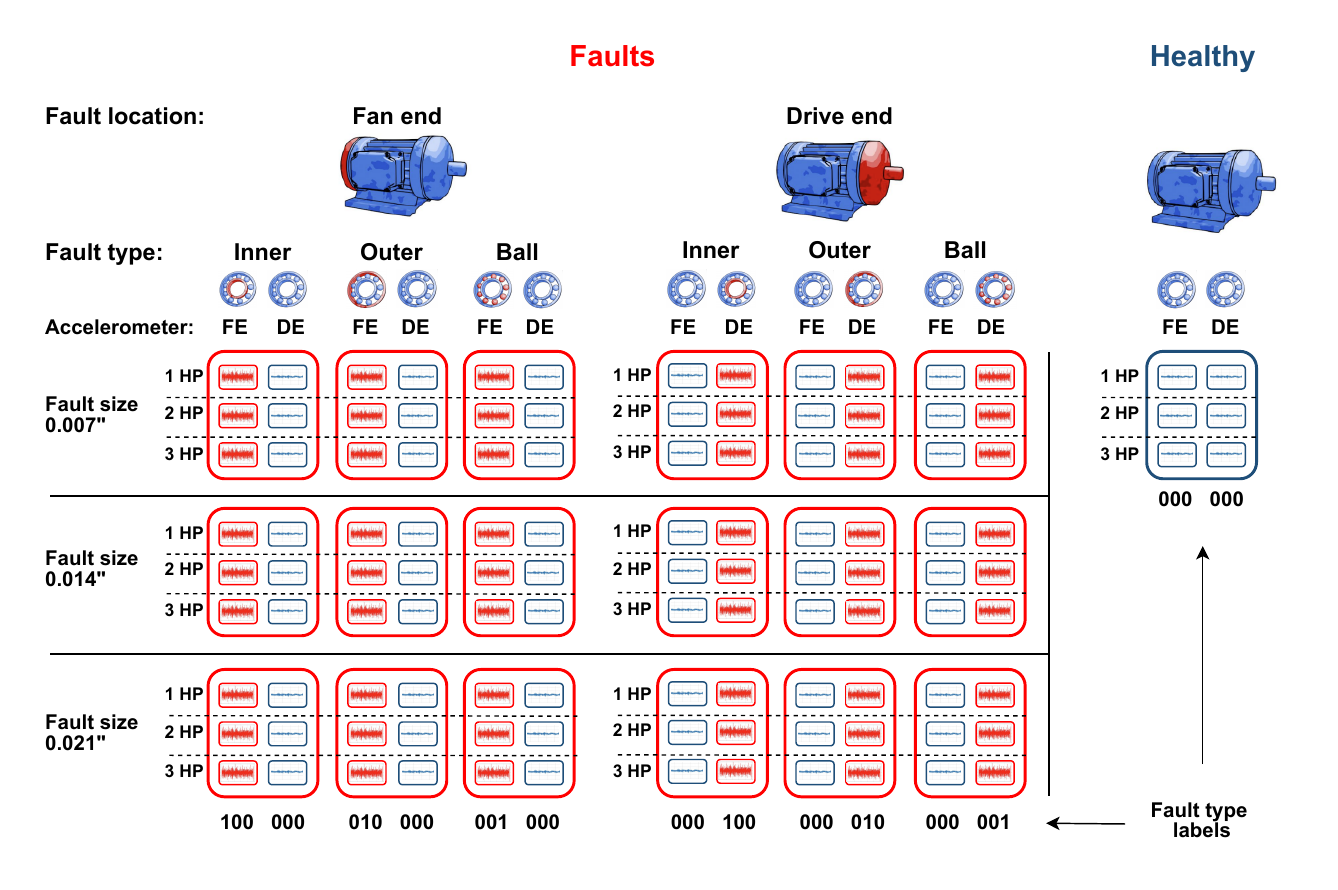}
    \caption{Signals from the CWRU bearing fault dataset considered in this work.
        Each box corresponds to a different bearing configuration (fault condition) specified by fault location, type and size, plus the healthy state. Six signals are acquired at two accelerometers (FE: fan end; DE: drive end) for each configuration at three different loads. The labels (which are the same for all signals in the same column) indicate the occurrence of an inner and/or outer and/or ball fault, in this order, at the same location where the signal is acquired.}
    \label{fig:cwru-configurations}
\end{figure*}

Following Hendriks et al. \cite{hendriks2022towards}, the configurations considered in this paper used a load varying from 1 to 3 HP and fault sizes of 7, 14 and 21 mils.
Measurements with a sampling rate of 12 kHz were used, except for the healthy bearing experiments that only had a sampling rate of 48 kHz available and were resampled to 12 kHz. Considering the three different fault positions at the outer race fault experiments, the ``Centered @6:00'' experiments were primarily used whenever possible. If the former did not exist, the ``Orthogonal @3:00'' experiments were used. All these configurations can be seen in Figure \ref{fig:cwru-configurations}, where each box represents a different bearing configuration. 
This paper also refers to these configurations as fault conditions (specified by fault location, type and size, plus the healthy state), as the dataset contains a single bearing configuration (a single pair of specific bearings) for each fault condition.

\subsection{Problem formulation}
\label{sec:methodology-multi-label}

The problem of detection and diagnosis of bearing faults can be formulated in many different ways, which can profoundly impact what a model can predict and how it is evaluated.

First, consider the problem of detecting faults located at the drive end using only signals acquired at the drive end. In this case, a simple binary healthy/faulty bearing classification is sufficient. However, suppose there is also a need to diagnose the fault type. If we assume that only a single fault type can happen simultaneously, i.e., the fault types are mutually exclusive, then multi-class classification is appropriate; the corresponding classes would be 
\textit{healthy}, \textit{inner}, \textit{outer} and \textit{ball}.
On the other hand, if we allow the possibility for multiple fault types to appear simultaneously, then (binary) multi-label classification should be used. In this case, each sample is labeled with three binary digits, 
each indicating whether an inner/outer/ball fault is present or not; in particular, the healthy state corresponds to the case where no fault type is present.

In a broader context,
our proposed multi-label approach to fault detection and diagnosis is that of constructing specific detectors for each possible type of fault. Then, conventional fault detection amounts to simply verifying if \textit{any} of the specific detectors gives a positive output, while fault diagnosis amounts to retrieving \textit{which} specific detectors give a positive output.

Besides being physically more realistic, this multi-label formulation presents several advantages in the case of the CWRU dataset, where fault types are mutually exclusive. First, the negative class samples for a given fault type, say inner, comprise samples with the other fault types, say outer and ball, besides the healthy samples. This mitigates the problem of the healthy class's heavy class imbalance (in lack of samples). More importantly, it allows us to move all the healthy samples to the test set, avoiding data leakage. Additionally, fine-grained, prevalence-independent evaluation metrics for binary classification, such as the ROC curve, can be used, resulting in a more precise evaluation.

Similar reasoning can be applied to the problem of identifying, additionally, the location of a fault (drive end or fan end) using both signals acquired at the drive end and the fan end.
Hendriks et al. \cite{hendriks2022towards} considered this problem under the assumption that the fault locations were mutually exclusive, formulating a multi-class classification problem with seven classes: 
\textit{healthy}, \textit{drive-inner}, \textit{drive-outer}, \textit{drive-ball}, \textit{fan-inner}, \textit{fan-outer} and \textit{fan-ball}.
Instead, we take a multi-label approach and treat the fault diagnosis at each location as independent problems. Specifically, each sample (consisting of a pair of DE/FE signals) is now labeled with two triples of binary digits, each triple containing the multi-label fault type classification for each location. Again, besides being more realistic, as it is physically possible for faults to arise at the two locations simultaneously, this formulation presents several advantages in the case of the CWRU dataset. 

First, as before, the samples with faults at the FE serve as negative samples for fault diagnosis at the DE and vice-versa. Second, since one signal is acquired at each location where a fault is to be diagnosed, we can split the problem in two. Namely, one model can be responsible for detecting faults at the DE based on signals acquired at the DE, while another model is analogous with respect to the FE. 
In this case, each DE or FE signal becomes a distinct sample and the labels are also split between the two signals: the DE signal is labeled with the triple of binary digits referring to faults at the DE, while the FE signal is labeled with the triple of binary digits referring to faults at the FE. In other words, each sample is labeled positively for a given fault type if and only if that fault type occurred at the specific location where the corresponding signal was acquired. Naturally, each DE or FE model is trained with only their respective DE or FE signals.
This approach has the additional advantage of relaxing the assumption that both signals are acquired synchronously, which is difficult to realize in practice. Indeed, this new formulation corresponds to the typical practical scenario where we wish to detect faults at a specific spot based on signals measured at that exact spot. 

Finally, we can exploit the symmetry of the problem (assuming that DE and FE bearings are sufficiently similar) and consider a single model to be used at both ends,
which is trained on all samples (DE and FE), effectively doubling the size of the training dataset. 
After a model is trained on such samples, the inference is made by running the model twice, once at each location, with the corresponding signal as input. This approach has the advantage of increasing the diversity of signals that a single model experiences, potentially improving its generalization to other locations.

The latter is our final approach to multi-label classification, summarized in Figure \ref{fig:cwru-configurations}.

\subsection{Dataset division}
\label{sec:dataset-division}

In the division proposed by Hendriks et al. \cite{hendriks2022towards}, faulty bearings with the same fault size are grouped, resulting in three different subsets (7, 14, 21 mils) that later are used to train and evaluate models.
For example, when the subset with a fault size of 7 mils is used for training, the remaining subsets of 14 and 21 mils are used as test sets.
This division prevents the occurrence of the same faulty bearings in both the train and test datasets, effectively addressing the data leakage issue. However, this solution is not entirely foolproof, 
as the dataset contains a single healthy bearing configuration, which must be split into the three previously described data subsets.
In the Hendriks et al. \citep{hendriks2022towards} approach, this division
is based on load, with healthy bearings at loads of 1, 2, and 3 HP being assigned to 
the 7, 14, and 21 mils subsets, respectively. It should be emphasized that any division of the healthy bearing configuration necessarily results in data leakage.

In contrast to the dataset division described above, in practice,
faults may occur in various sizes, making it unrealistic to have a dataset distribution of only one fault size during training. This can lead to difficulty in training a model to detect faults at different sizes that did not appear in the training set. Therefore, it is important to consider a more diverse range of fault conditions in the dataset to ensure that the model can accurately predict them.

A more realistic approach would be a division where multiple loads, fault sizes, types, and locations randomly appear in each subset. Hence, the proposed division selects signals from a random fault size configuration for each fault location and type pair, provided that signals from the same bearing configuration appear only in a single (train or test) set to avoid data leakage. The selected signals---together with all the healthy bearing signals---are used for the test set, and the remaining signals are used in the training set. This results in $3^{6}$ possible train-test splits.

In addition to this hold-out method, it is also possible to conduct 3-fold cross-validation using the proposed division. The method begins by randomly selecting fault size configurations for each fault location and type pair, mirroring the process of the hold-out method. However, instead of designating this sample set as a test set, it becomes our initial fold. To generate the remaining two folds, a subsequent hold-out split is executed on the remaining configurations after excluding the first fold. This leads to a total of $3^6\times2^6/6$ potential 3-fold cross-validation splits. Note that only the faulty bearing signals are considered in this process, with the healthy bearing signals always being placed in each corresponding test set.

\begin{figure*}[t]
    \centering
    \includegraphics[scale=0.45]{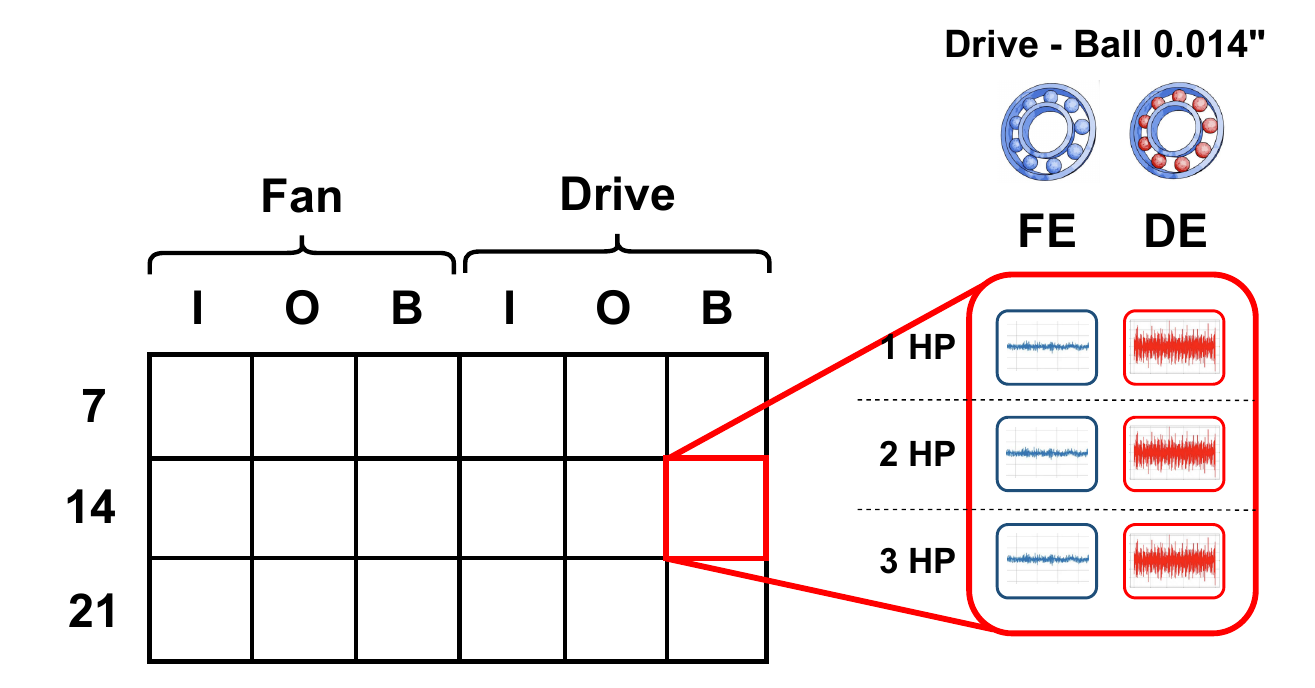}
    \caption{Simplified dataset representation. Each column represents a fault type (I: Inner race; O: Outer race; B: Ball) and location, and each row represents a fault size. Each cell represents all the signals acquired in the corresponding fault condition (which, in the CWRU dataset, comprises a single-bearing configuration). This representation emphasizes that all signals acquired in the same bearing configuration, being potentially correlated, should belong to the same (train or test) data subset.}
    \label{fig:diagram-explanation.pdf}
\end{figure*}

\begin{figure*}[t]
    \centering
    \begin{subfigure}{0.45\textwidth}
        \centering
        \includegraphics[width=\linewidth]{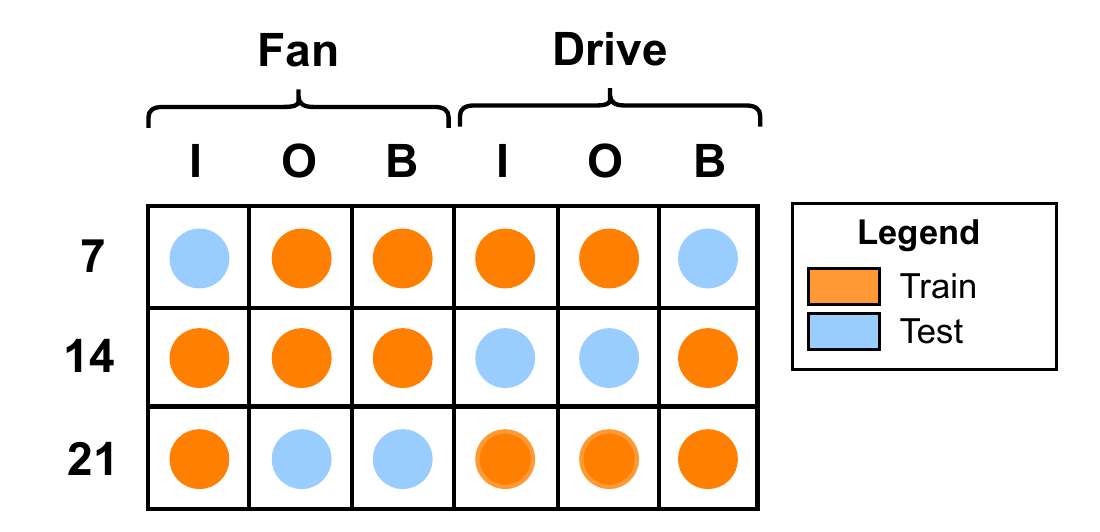}
        \caption{Example of a train-test split.} 
        \label{fig:train_test_division_example}
    \end{subfigure}
    \hspace{10pt}
    \begin{subfigure}{0.45\textwidth}
        \centering
        \includegraphics[width=\linewidth]{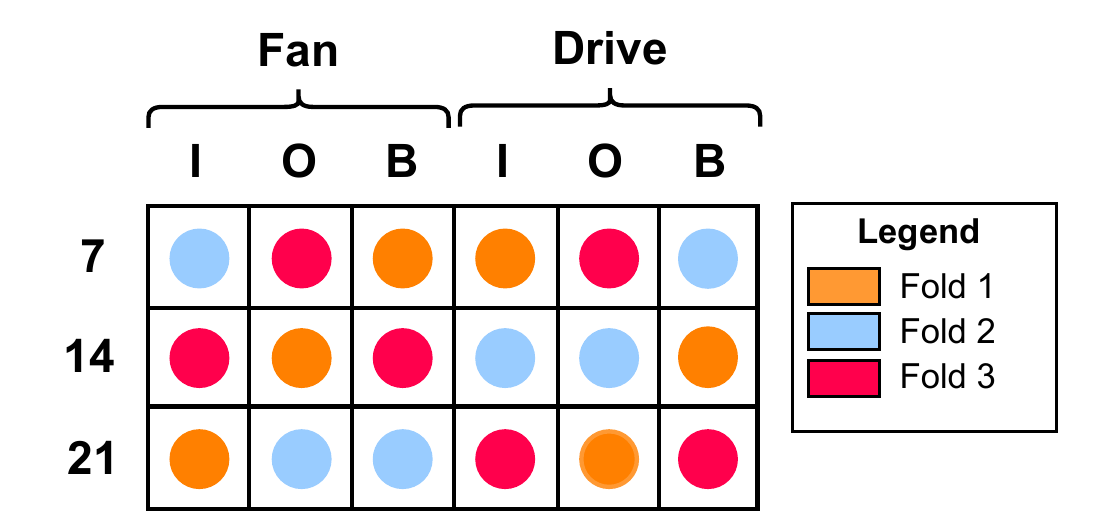}
        \caption{Example of a 3-fold partition.}
        \label{fig:three_fold_division_example1}
    \end{subfigure}
    \caption{Example of proposed division. Each column must contain exactly a single test group (blue dot). Note that all healthy signals are always placed in the test set.} 
    \label{fig:division-diagram.pdf}
\end{figure*}

Figure \ref{fig:diagram-explanation.pdf} provides a simplified representation of the fault samples in the dataset shown in Figure \ref{fig:cwru-configurations}. Each cell in Figure \ref{fig:diagram-explanation.pdf} corresponds to a fault condition (specified by a fault location/type column and a fault size row) and represents all the signals acquired in the corresponding configuration (namely, the signals acquired by both fan end and drive end accelerometers for every single load).
Figure~\ref{fig:division-diagram.pdf} shows an example of a possible train-test split for the fault signals, as well as an example of a 3-fold partition. It is important to note that, while not represented in Figure~\ref{fig:division-diagram.pdf}, the 6 signals corresponding to healthy bearings are always present in the test set.

\subsection{Model selection and evaluation methodology}
\label{sec:methodology-hparam}
For hyperparameter optimization (also called model selection) and performance evaluation, we adopt the Double Cross-Validation Method (CVM-CV), described in \cite{tsamardinos2015performance}. This protocol consists of applying the cross-validation method for hyperparameter optimization (CVM) and then reevaluating it on different train-test splits (CV) for final performance estimation for the single, selected, best model. Note that using only CVM is well-known to overestimate performance since it returns the maximum performance achieved across several hyperparameter configurations. This bias can be reduced by reevaluating only the selected hyperparameter configuration on different train-test splits.

In this paper, the CVM-CV method is used on the CWRU bearing fault dataset by applying a CVM step on a random 3-fold partition as described in Section \ref{sec:dataset-division}, followed by a CV step using 30 random train-test splits (with static seeds) to evaluate performance. The seeds are reused across experiments to ensure that the 
different models are tested on exactly the same test sets.

\subsection{Model architectures and training details}
\label{sec:methodology-model}

Numerous techniques have been proposed for bearing fault diagnosis, including signal processing \cite{randall2021vibration, smith2015rolling}, machine learning \cite{lei2020applications} and deep learning \cite{jia2016deep}. Among these techniques, Convolutional Neural Networks (CNNs) have gained wide attention and exploration due to their exceptional performance and efficiency in image recognition systems. These models can be used for various tasks, including fault diagnosis, and have shown remarkable results  \cite{zhang2019deep, zhang2017new, zhang2018deep}. They typically use 1D or 2D vibration signal representations on time, frequency, or time-frequency domains and can be trained either from scratch or fine-tuned from pre-trained image models.

In this paper, the ResNet architecture \cite{he2016deep} is used as the primary model for the experiments due to its promising results in the literature. 
The batch size, learning rate, and the number of epochs were optimized, aiming to maximize macro average AUROC by using a grid search strategy along the CVM-CV method described in \ref{sec:methodology-hparam}. The hyperparameter grid included batches of sizes 32, 64, and 128, with learning rates between $10^{-7}$ and $10^{-3}$ (with a multiplicative step of 10). Experiments were done with 10 training epochs but with a checkpointing to save only the epoch with the highest metric.
After hyperparameter optimization, the ResNet18 architecture was fine-tuned for four epochs, using a batch size of 128, a constant learning rate of $10^{-5}$ and the Adam optimizer \cite{kingma2014adam} starting from weights pre-trained on the ImageNet dataset.

Given the results with a pre-trained ImageNet vision model, new experiments were proposed with other similar model architectures from PyTorch's torchvision models \cite{torchvision2016} such as MobileNet  \cite{howard2019searching}, RegNet \cite{radosavovic2020designing}, ConvNeXt \cite{liu2022convnet}, Swin Transformer \cite{liu2022swin}, ViT \cite{dosovitskiy2020image}, EfficientNet \cite{tan2021efficientnetv2} and MaxVit \cite{tu2022maxvit}. As simpler baseline models, the 7-layer CNN and 9-layer CNN \cite{zhang2021cjc} (typically used in experiments on the CIFAR 10/100 dataset) were also included in our experiments. We also experimented with different-sized ResNet architectures \cite{he2016deep}, such as the ResNet34, ResNet50, and ResNet101. 
All the models used pre-trained model weights and binary cross-entropy loss. For the experiments with 1D signal representations on the time or frequency domain, we utilized the WDCNN model architecture \cite{zhang2017new}. The WDCNN is a 1D CNN that was built with the purpose of being used on machine fault diagnosis problems and has shown results close to those obtained by the ResNet model \cite{hendriks2022towards}.
All of the model's hyperparameters, such as batch size, learning rate, and number of epochs, were optimized using exactly the same procedure as for the ResNet18 architecture, the optimized hyperparameter can be seen in Table \ref{tab:hparams}.

\begin{table}[t]
\centering
\caption{Tuned hyperparameters for all models.}
\label{tab:hparams}
{
\fontsize{9pt}{11pt}\selectfont
\tabcolsep5pt
\renewcommand{\arraystretch}{1.2}
\begin{tabular}{@{}llcccc@{}}
\toprule
\multicolumn{2}{l}{\multirow{2.3}{*}{\textbf{Model}}} & \multicolumn{1}{l}{\multirow{2.3}{*}{\textbf{\# Param.}}} & \multicolumn{3}{c}{\textbf{Hyperparameters}} \\ \cmidrule(l){4-6} 
\multicolumn{2}{l}{}                                & \multicolumn{1}{l}{}                                 & \textbf{Batch Size} & \textbf{Learning Rate} & \textbf{Epoch} \\ \midrule
\multicolumn{2}{l}{\textbf{ResNet18 \cite{he2016deep}}}        & \textbf{11.7M} & \textbf{128} & \bm{$10^{-5}$} & \textbf{4} \\ \midrule
\multicolumn{2}{l}{ResNet34 \cite{he2016deep}}                 & 21.8M & 128 & $10^{-5}$ & 3 \\
\multicolumn{2}{l}{ResNet50 \cite{he2016deep}}                 & 25.6M & 128 & $10^{-6}$ & 7 \\
\multicolumn{2}{l}{ResNet101 \cite{he2016deep}}                & 44.5M & 128 & $10^{-6}$ & 9 \\ \midrule
\multicolumn{2}{l}{CNN7 \cite{zhang2021cjc}}                     & 40M   & 32  & $10^{-5}$ & 1 \\
\multicolumn{2}{l}{CNN9 \cite{zhang2021cjc}}                     & 2.8M  & 64  & $10^{-3}$ & 5 \\
\multicolumn{2}{l}{MobileNet V3 (large) \cite{howard2019searching}}     & 5.5M  & 32  & $10^{-4}$ & 7 \\
\multicolumn{2}{l}{RegNet (x\_1\_6gf) \cite{radosavovic2020designing}}       & 9.2M  & 64  & $10^{-4}$ & 3 \\ \midrule
\multicolumn{2}{l}{ConvNeXt (tiny) \cite{liu2022convnet}}          & 28.6M & 64  & $10^{-4}$ & 1 \\
\multicolumn{2}{l}{SwinTransformerV2 (tiny) \cite{liu2022swin}} & 28.4M & 64  & $10^{-4}$ & 3 \\
\multicolumn{2}{l}{ViT (b\_16) \cite{dosovitskiy2020image}}               & 86.6M & 64  & $10^{-6}$ & 1 \\
\multicolumn{2}{l}{EfficientNetV2 (small) \cite{tan2021efficientnetv2}}   & 21.5M & 64  & $10^{-4}$ & 1 \\
\multicolumn{2}{l}{MaxViT \cite{tu2022maxvit}}                   & 30.9M & 32  & $10^{-5}$ & 1 \\ \midrule
\multicolumn{2}{l}{WDCNN \cite{zhang2017new} - Time} & 53.5K & 32 & $10^{-4}$ & 6 \\
\multicolumn{2}{l}{WDCNN \cite{zhang2017new} - Spectrum} & 53.5K & 32 & $10^{-3}$ & 8 \\
\multicolumn{2}{l}{WDCNN \cite{zhang2017new} - Power Cepstrum} & 40.7K & 32 & $10^{-3}$ & 6 \\
\bottomrule
\end{tabular}%
}
\end{table}

The pre-processing steps followed those in \cite{hendriks2022towards}. For the 2D models, it involved in a segmentation with an overlap of 97\% (resulting in segments with a window length of 11500) and calculating the spectrogram of each segment 
using a window size of 104 points, 54 points of overlap, and 452 DFT points. These parameters give each resulting spectrogram a size of 227x228. However, to match the ResNet model's 224x224 network input size, the spectrogram is cropped by keeping the lower-left portion (removing some of the high frequencies and some of the last few time steps).

For the 1D models, three different signal representations were used (time, spectrum and power cepstrum) and the pre-processing steps differed slightly for each of them. For the time-domain signal, raw temporal accelerometer data is segmented with the same overlap of 97\% but with a window length of 2048 samples. For the spectrum and power cepstrum signal representations, a window length of 4096 samples is used. The spectrum is computed by taking the Fast Fourier Transform (FFT) of the raw data, while the power cepstrum is computed by taking the power spectrum of the logarithm of the signal's power spectrum.

The 1D and 2D features were then scaled using z-score normalization fitted on the whole training set (i.e., resulting in a single scalar for the mean across all features and all inputs and similarly for the standard deviation).


\section{Experiments and results}
\label{sec:experiments}

The experiments were conducted using Python with the PyTorch framework on a machine equipped with 
an RTX 3090 GPU. For each model, after hyperparameter optimization, each train-test run was repeated 30 times with different random dataset divisions as explained in Section \ref{sec:methodology-hparam}, and the final results presented are the averages and standard deviations of those 30 runs.  

The results and error analysis of the proposed methodology are given in Section \ref{subsec:results}, with ablation experiments for the proposed method being presented in Section \ref{subsec:ablation-studies}. Further experiments on different convolutional vision models pre-trained on ImageNet are given in Section \ref{subsec:different-architectures}. 

\subsection{Multi-label fault diagnosis} 
\label{subsec:results}

\begin{figure*}[!ht]
    \centering
    \begin{subfigure}{0.45\textwidth}
        \centering
        \includegraphics[width=\linewidth]{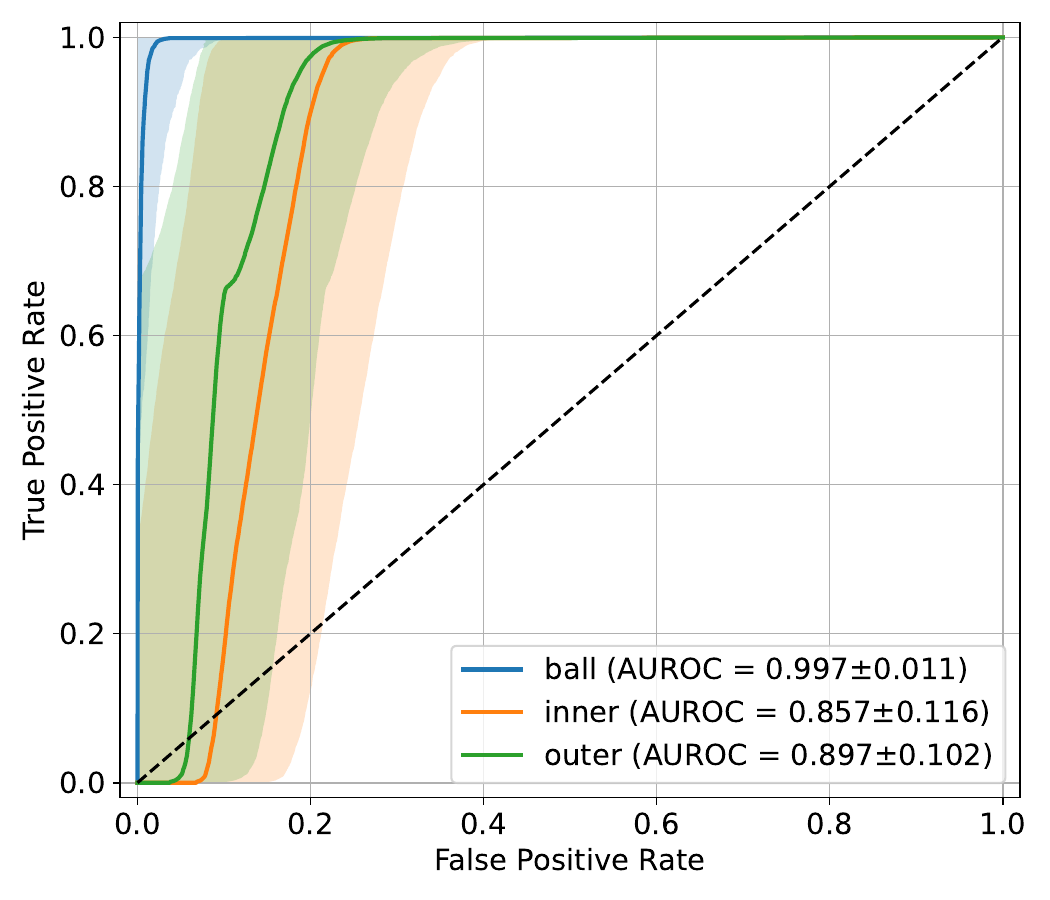}
        \caption{Fan end} 
        \label{fig:roc_curve_by_location_a}
    \end{subfigure}
    \hspace{15pt}
    \begin{subfigure}{0.45\textwidth}
        \centering
        \includegraphics[width=\linewidth]{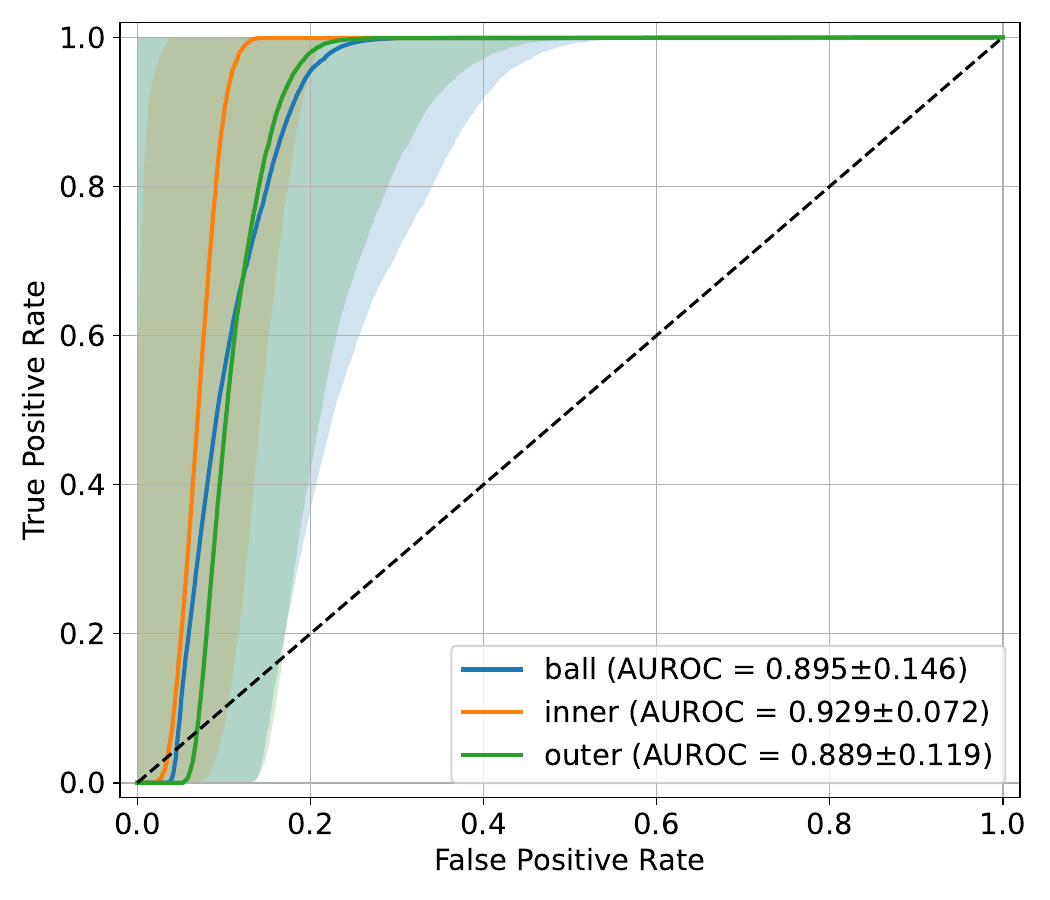}
        \caption{Drive end}
        \label{fig:roc_curve_by_location_b}
    \end{subfigure}
    \vspace{\floatsep}
    \begin{subfigure}{1\textwidth}
        \centering
        \includegraphics[scale=0.45]{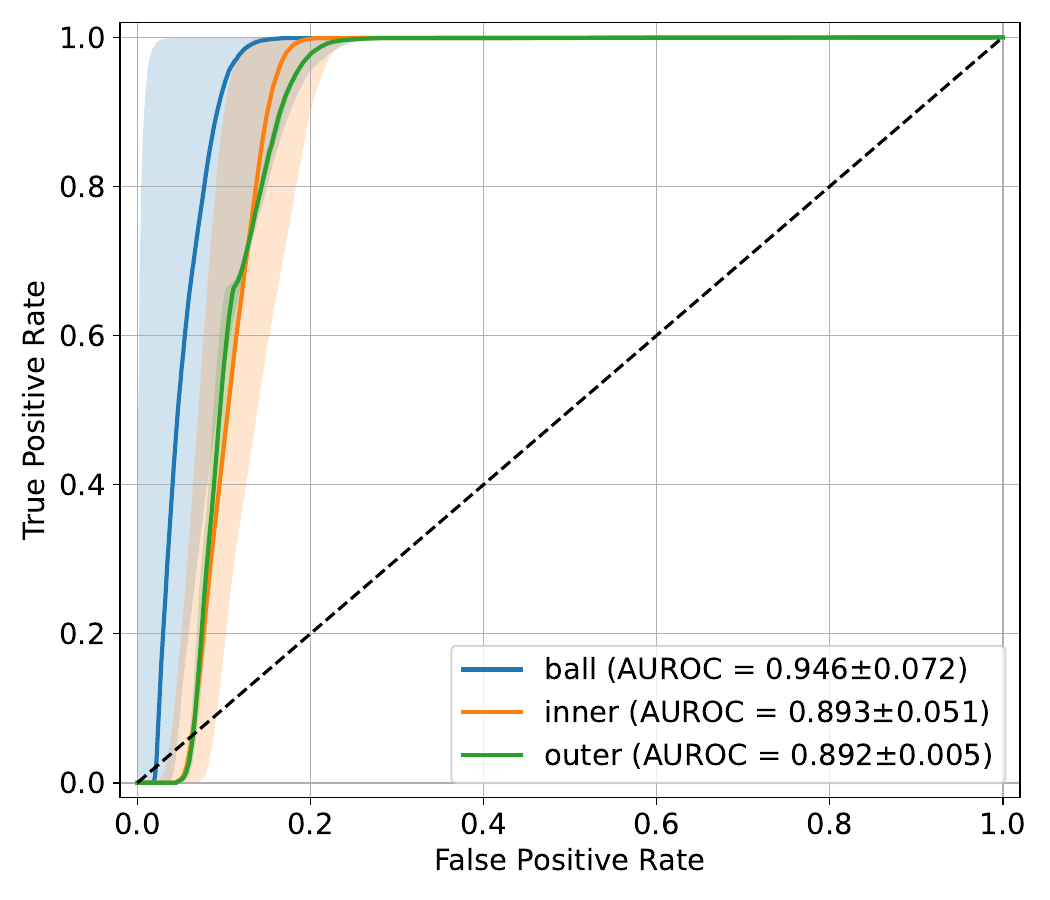}
        \caption{Horizontal average of (a) and (b)}
    \label{fig:roc_curve_average}
    \end{subfigure}
    
    \caption{ROC curves for multi-label fault diagnosis at each location. For (a) and (b), solid curves represent the horizontal average ROC curve across 30 realizations, while the solid curves in (c) represent the average across all realizations from (a) and (b). In all cases, the filled region represents the standard deviation.} 
    \label{fig:roc_curve}
\end{figure*}

Table \ref{tab:result} shows the AUROC of the ResNet18 model for each fault location/type, while Figure \ref{fig:roc_curve} shows the corresponding ROC curves. 
Note that while we have used a single model to diagnose faults at both locations, in principle, each location defines a separate problem and, therefore, a potentially different detector (in particular, we may use different decision thresholds for each fault location/type detector). Thus, the results are presented separately for each fault location/type and subsequently averaged.

As can be seen, all detectors show good performance, with a macro-average AUROC of $0.911\pm0.044$. The highest and lowest performance are achieved, respectively, by the ball fault detector at the fan end, with an AUROC of $0.997\pm0.011$, and the inner fault detector at the fan end, with an AUROC of $0.857\pm0.114$.

This difference in performance, and in particular in the standard deviation for each ROC curve, 
can be explained by the error analysis in the next subsection.

Since false negatives usually have worse consequences in practice than false positives in the case of fault detection and diagnosis, fixing the value of TPR (which is the complement of the false negative rate) is an appropriate way of setting an acceptable fault tolerance for false negatives, after which the FPR can be measured and compared. Assuming that 90\% TPR is an acceptable rate for the system and considering the average ROC curve of Figure \ref{fig:roc_curve_average}, the resulting fault-type FPRs (averaged across location) are 9.2\%, 15\%, and 16.5\% for the ball, inner, and outer fault types, respectively.

\begin{table}[ht]
\centering
\caption{100 $\times$ AUROC (mean $\pm$ std) for the ResNet18 model.}
\label{tab:result}
{
\fontsize{9pt}{11pt}\selectfont

\tabcolsep4pt
\begin{tabular}{@{}llllllllll@{}}
\toprule
  \multicolumn{3}{c}{\textbf{Fan end}} &
  \multicolumn{3}{c}{\textbf{Drive end}} &
  \multicolumn{3}{c}{\textbf{FE/DE Average}} &
  \multirow{2.3}{*}{\begin{tabular}[c]{@{}c@{}}\textbf{Macro} \\ \textbf{Average}\end{tabular}} \\ \cmidrule(lr){1-3} \cmidrule(lr){4-6} \cmidrule(lr){7-9} 
  \multicolumn{1}{c}{\textbf{Ball}} &
  \multicolumn{1}{c}{\textbf{Inner}} &
  \multicolumn{1}{c}{\textbf{Outer}} &
  \multicolumn{1}{c}{\textbf{Ball}} &
  \multicolumn{1}{c}{\textbf{Inner}} &
  \multicolumn{1}{c}{\textbf{Outer}} &
  \multicolumn{1}{c}{\textbf{Ball}} &
  \multicolumn{1}{c}{\textbf{Inner}} &
  \multicolumn{1}{c}{\textbf{Outer}} &
   \\[4pt] \midrule
  99.7 \scriptsize$\pm$1.1 &
  85.7 \scriptsize$\pm$11.6 &
  89.7 \scriptsize$\pm$10.2 &
  89.5 \scriptsize$\pm$14.6 &
  92.9 \scriptsize$\pm$7.2 &
  88.9 \scriptsize$\pm$11.9 &
  94.6 \scriptsize$\pm$7.2&
  89.3\scriptsize$\pm$5.1&
  89.2\scriptsize$\pm$0.5&
  91.1 \scriptsize$\pm$4.4 \\[4pt]\bottomrule
\end{tabular}%
}
\end{table}

\subsubsection{Error analysis}
\label{subsubsec:error-analysis}

\begin{figure*}[t]
    \centering
    \includegraphics[width=\linewidth]{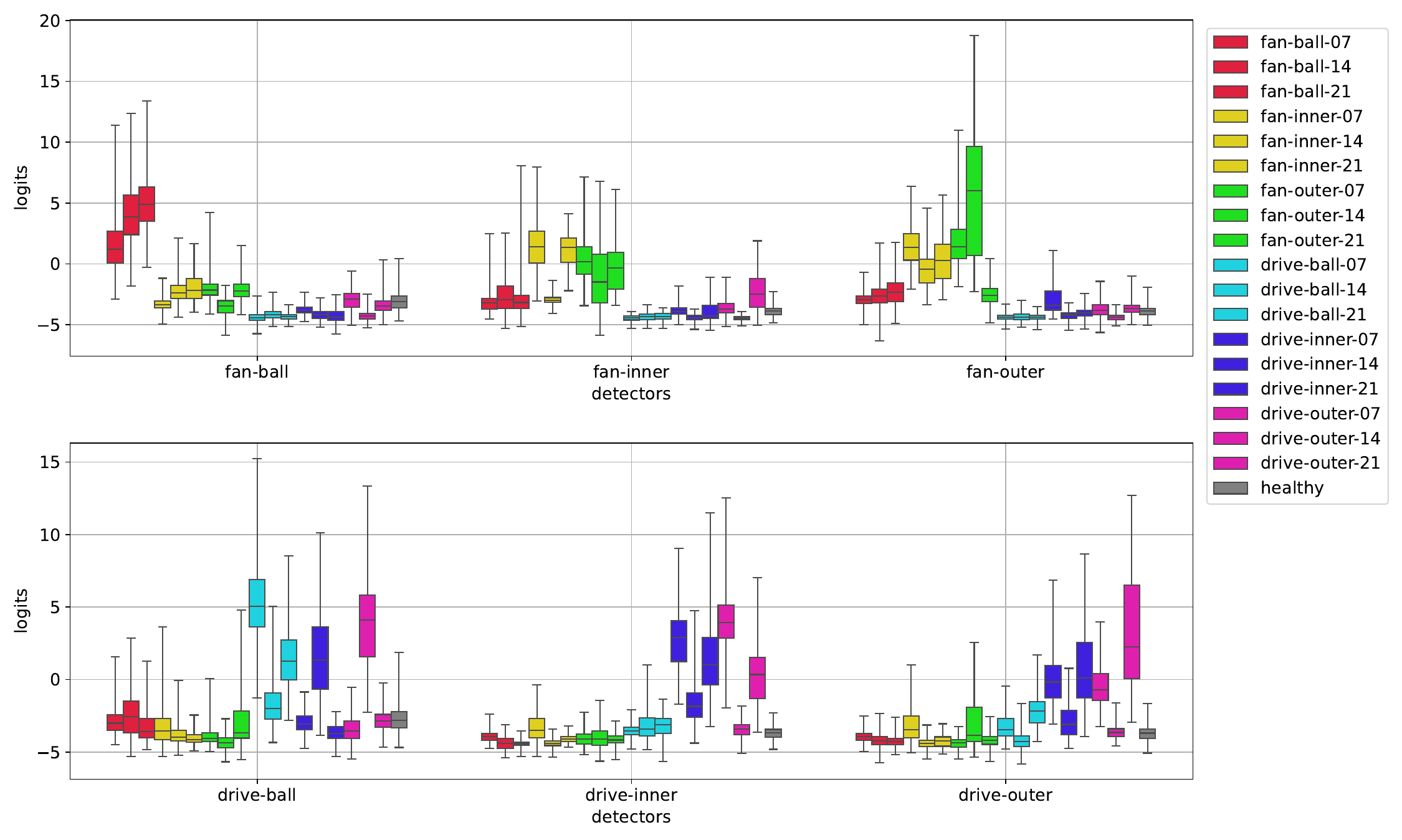}
    \caption{Boxplots of the raw confidence scores (logits) for fault diagnosis. Each group of boxplots, identified by the colors, corresponds to the detection of a specific fault type at a specific location, and the fault sizes are ordered by size ranging from 7 to 21 mils. Each boxplot within a group corresponds to one of 19 fault conditions (denoted by fault location-type-size plus the healthy state) and displays the spread of logits among all train-test realizations where that condition appeared in the test set. At each boxplot, the whiskers represent the minimum and maximum values achieved.}
    \label{fig:logits-boxplots}
\end{figure*}

Figure~\ref{fig:logits-boxplots} shows boxplot visualizations of the output \textit{logits} (raw confidence scores for the positive class, before squashing by the logistic function), for each fault location/type detector, obtained by grouping, for each of the 19 fault conditions, all realizations where that condition appears in the test set.
The higher the logit output, the higher the confidence of the model that there is a fault of that specific type at that specific location. Usually, multi-label models have an added step of passing this score through a sigmoid activation function to transform it into a probability between 0 and 1, but this step was removed for this visualization to allow better interpretation without the probability capping at 1.

As can be seen in Figure~\ref{fig:logits-boxplots}, most detectors prioritize the correct fault conditions, 
by assigning higher scores than those of
the corresponding negative conditions.
(Note that only relative values matter, as different thresholds can be set for each detector in order to reach a desired trade-off between true and false positives.)
However, with the exception of the fan-ball detector, which reached almost perfect discrimination, there was often some confusion between fault types at the same location, mostly between inner and outer faults.
In particular, the fault conditions fan-inner-14, fan-outer-21, drive-inner-14, and drive-outer-14 achieved scores lower than expected for their respective detectors, while the latter two conditions additionally achieved (incorrectly) high scores on the drive-ball detector.

These errors can be partly explained with the help of the comprehensive analysis of the CWRU dataset performed by Smith and Randall \cite{smith2015rolling}. They applied three established diagnostic techniques and found that relatively few of the records gave classical characteristics of the fault types, with many of the records showing evidence of mechanical looseness. As a consequence, some records were cited as difficult to diagnose or not diagnosable with any of the applied methods; these include, among the signals considered in our work, the drive-inner-14 and drive-outer-14 conditions and most of the ball faults. Note that, according to our dataset division, when a particular fault condition (say, drive-inner-14) is included in the test set, the other two fault sizes of the same condition (drive-inner-07 and drive-inner-21) necessarily appear in the training set. If only the former is difficult to diagnose, it likely will not share similar characteristics with the other ones in the training set, and therefore, its scores will be low for the corresponding detector. Conversely, when difficult-to-diagnose (and possibly anomalous) signals appear in the training set, the model will be forced to learn unconventional (and possibly spurious) features of these signals in order to push their scores up, which may inadvertently give false positives to other fault types. This helps explain the frequent confusion between inner and outer faults, as well as why precisely the drive-inner-14 and drive-outer-14 conditions achieve a high score on the drive-ball detector (they likely share spurious features with the ball fault signals at the same location).

Nevertheless, it becomes clear from Figure~\ref{fig:logits-boxplots} that the model rarely mistakes the fault location, a property which is further explored in Section~\ref{subsec:fault-detection}.

Another interesting observation that can be made from Figure~\ref{fig:logits-boxplots} is that there is apparently no correlation between fault score and fault size. One might assume that fault scores would be higher for larger fault sizes, but that is not the case. Again, this has been previously suggested by Smith and Randall in \cite{smith2015rolling}, who observed that the fault size seemed to have a lower impact on the diagnosis outcomes than the test rig assembly (since disassembly and reassembly are required to replace a bearing with a different fault size). These results suggest that it may be impossible to predict fault sizes using this dataset.

\subsection{Ablation studies}
\label{subsec:ablation-studies}

Several experiments were performed to evaluate the proposed methodology on different choices regarding model specificity, train-test split type and proportion, and segment length. The first ablation study investigated what was most advantageous, having a separate model for each bearing location or a single model trained on data from every bearing and evaluated once at each location (the latter is our proposed approach). In principle, using separate models makes sense since each bearing type (and its external environment) induces a potentially different data distribution, formally known as a \textit{domain}, so each model could benefit from specializing to a single domain. On the other hand, we should expect the single model approach to perform better if the two domains are not too different since the model would then be exposed to twice the amount of training data.

In Table~\ref{tab:ablations}, the first row represents the single model approach following the methodology described in Section~\ref{sec:methodology-multi-label}, while the second row considers a separate model for the DE and FE locations. 
We can see a drop in performance for the separate approach, indicating that learning from these two domains jointly is viable and beneficial.


\begin{table}[ht]
\centering
\caption{
100 $\times$ AUROC (mean $\pm$ std) for ablation experiments using the ResNet18 architecture. Numbers in boldface indicate the best result for each column.}
\label{tab:ablations}
{
\fontsize{9pt}{11pt}\selectfont

\tabcolsep4pt
\begin{tabular}{@{}lllllllllll@{}}
\toprule
\multirow{2.3}{*}{\textbf{Model}} &
  \multirow{2.3}{*}{\textbf{Split type}} &
  \multirow{2.3}{*}{\begin{tabular}[c]{@{}l@{}}\textbf{Split} \\ \textbf{ratio}\end{tabular}} &
  \multirow{2.3}{*}{\begin{tabular}[c]{@{}l@{}}\textbf{Signal} \\ \textbf{length}\end{tabular}} &
  \multicolumn{3}{c}{\textbf{Fan}} &
  \multicolumn{3}{c}{\textbf{Drive}} &
  \multirow{2.3}{*}{\begin{tabular}[c]{@{}c@{}}\textbf{Macro} \\ \textbf{Average}\end{tabular}} \\ \cmidrule(lr){5-7} \cmidrule(lr){8-10}
 &
   &
   &
   &
  \multicolumn{1}{c}{\textbf{Ball}} &
  \multicolumn{1}{c}{\textbf{Inner}} &
  \multicolumn{1}{c}{\textbf{Outer}} &
  \multicolumn{1}{c}{\textbf{Ball}} &
  \multicolumn{1}{c}{\textbf{Inner}} &
  \multicolumn{1}{c}{\textbf{Outer}} &
   \\[4pt] \midrule
Single &
  Proposed &
  2:1 &
  Full &
  \textbf{99.7 \scriptsize$\pm$1.1} &
  \textbf{85.7 \scriptsize$\pm$11.4} &
  \textbf{89.7 \scriptsize$\pm$10.0} &
  89.5 \scriptsize$\pm$14.3 &
  \textbf{92.9 \scriptsize$\pm$7.1} &
  \textbf{88.9 \scriptsize$\pm$11.7} &
  \textbf{91.1 \scriptsize$\pm$4.4} \\[4pt]
\multirow{2}{*}{\begin{tabular}[c]{@{}l@{}}Separate\\[-1pt] DE/FE\end{tabular}} &
  \multirow{2}{*}{Proposed} &
  \multirow{2}{*}{2:1} &
  \multirow{2}{*}{Full} &
  \multirow{2}{*}{99.1 \scriptsize$\pm$3.0} &
  \multirow{2}{*}{82.7 \scriptsize$\pm$12.0} &
  \multirow{2}{*}{89.5 \scriptsize$\pm$10.7} &
  \multirow{2}{*}{90.6 \scriptsize$\pm$13.7} &
  \multirow{2}{*}{92.0 \scriptsize$\pm$7.5} &
  \multirow{2}{*}{76.5 \scriptsize$\pm$13.3} &
  \multirow{2}{*}{88.4 \scriptsize$\pm$7.2} \\
 &
   &
   &
   &
   &
   &
   &
   &
   &
   &
   \\[4pt]
Single &
By fault size
  &
  2:1 &
  Full &
  94.5 \scriptsize$\pm$2.4 &
  80.5 \scriptsize$\pm$4.1 &
  88.9 \scriptsize$\pm$1.3 &
  81.5 \scriptsize$\pm$21.7 &
  84.3 \scriptsize$\pm$5.5 &
  76.6 \scriptsize$\pm$3.2 &
  84.4 \scriptsize$\pm$5.9 \\[4pt]
Single &
  Proposed &
  1:2 &
  Full &
  70.0 \scriptsize$\pm$14.1 &
  84.6 \scriptsize$\pm$11.3 &
  73.1 \scriptsize$\pm$17.8 &
  92.5 \scriptsize$\pm$6.9 &
  77.9 \scriptsize$\pm$7.5 &
  80.3 \scriptsize$\pm$15.4 &
  79.8 \scriptsize$\pm$7.4 \\[4pt]
Single &
  Proposed &
  2:1 &
  Half &
  98.9 \scriptsize$\pm$3.2 &
  83.4 \scriptsize$\pm$13.8 &
  89.0 \scriptsize$\pm$11.6 &
  \textbf{90.7 \scriptsize$\pm$13.1} &
  92.6 \scriptsize$\pm$7.7 &
  88.1 \scriptsize$\pm$14.1 &
  90.4 \scriptsize$\pm$4.7 \\[4pt]
\multirow{2}{*}{\begin{tabular}[c]{@{}l@{}}Separate\\ DE/FE\end{tabular}} &
  \multirow{2}{*}{By fault size} &
  \multirow{2}{*}{1:2} &
  \multirow{2}{*}{Full} &
  \multirow{2}{*}{89.0 \scriptsize$\pm$1.0} &
  \multirow{2}{*}{76.7 \scriptsize$\pm$2.8} &
  \multirow{2}{*}{73.6 \scriptsize$\pm$3.5} &
  \multirow{2}{*}{82.8 \scriptsize$\pm$2.7} &
  \multirow{2}{*}{90.6 \scriptsize$\pm$2.9} &
  \multirow{2}{*}{56.0 \scriptsize$\pm$4.5} &
  \multirow{2}{*}{78.1 \scriptsize$\pm$11.6} \\
 &
   &
   &
   &
   &
   &
   &
   &
   &
   &
   \\[4pt]\bottomrule
\end{tabular}%
}
\end{table}

Another ablation considered splitting the data into train and test sets
based on fault size. In this approach, the bearing configurations for one of the fault sizes (7, 14, 21 mils) were allocated to the test set, along with the healthy state, while configurations with other fault sizes were used for training. This is similar to what is done in \cite{hendriks2022towards}, but with a train/test proportion of 2:1 instead of their 1:2 proportion. This results in three different train/test splits that are repeated 10 times for each split, resulting in 30 runs, just like the other experiments.
The macro average AUROC dropped from 0.911 to 0.844, showing that the increase in data diversity (without leakage) brought by the proposed random split of Section~\ref{sec:dataset-division} is indeed beneficial. 

Many papers consider a 1:2 train/test split proportion when generating subsets based on load or fault size. From a purely machine learning perspective, this can be seen as an unconventional choice, as it is common practice to reserve a larger fraction of the data for training, arguably because models with sufficient capacity can benefit from more training data. To test whether this is the case for our problem, an experiment was performed using the 1:2 proportion, and as can be seen in Table~\ref{tab:ablations}, a considerable drop in performance was observed. This result, in turn, raises another question: is this loss in performance due to merely a smaller training set or due to a less \textit{diverse} training set? To answer this question, 
an ablation experiment was performed using only the first half of each signal before segmentation, thus resulting in the same training set size as the 1:2 split but with the same diversity of fault conditions as the 2:1 split. The results show a macro average AUROC of 0.904, which is quite close to that of the original model, indicating that the leading cause of the drop in performance is the reduced diversity in the training data---or, equivalently, that increased training data diversity is highly beneficial. 

A final experiment was made considering a baseline training approach with separate models for each location and a train/test split by fault size with a 1:2 split ratio (an attempt to approximate the setup of \cite{hendriks2022towards} under our formulation), which resulted in a macro average AUROC of 0.781. It follows that the overall effect of our training approach is to significantly improve the performance by about 13 AUROC percentage points over the baseline.

\subsection{Other model architectures}
\label{subsec:different-architectures}

In order to select the most suitable model architecture, various experiments were conducted using different convolutional vision models that were pre-trained on the ImageNet dataset, except for the CNN7 and CNN9 model architectures, which were trained from scratch and used as baselines. The experiments were divided into models that were simpler/smaller and more complex/larger than the ResNet18 to determine if any conclusions could be drawn regarding model complexity and performance. 
It is important to note that the hyperparameters were optimized (and subsequently evaluated) for each experiment using the same model selection and evaluation methodology described in Section \ref{sec:methodology-hparam}.

In Tables \ref{tab:smaller-models} and \ref{tab:larger-models}, it can be observed that neither the smaller nor larger models evaluated in these experiments yielded a better result than what was previously obtained with the ResNet18, with the smaller models MobileNet and RegNet having a slight edge over the larger models present in Table \ref{tab:larger-models}.

\begin{table}[ht]
\centering
\caption{100 $\times$ AUROC (mean $\pm$ std) for simpler/smaller model architectures. Numbers in boldface indicate the best result for each column.}
\label{tab:smaller-models}
{
\fontsize{9pt}{11pt}\selectfont
\tabcolsep5pt
\renewcommand{\arraystretch}{1.2}
\begin{tabular}{@{}lllllllll@{}}
\toprule
\multicolumn{2}{l}{\multirow{2.3}{*}{\textbf{Model}}} &
  \multicolumn{3}{c}{\textbf{Fan}} &
  \multicolumn{3}{c}{\textbf{Drive}} &
  \multirow{2.3}{*}{\begin{tabular}[c]{@{}c@{}}\textbf{Macro} \\ \textbf{Average}\end{tabular}} \\ \cmidrule(lr){3-5} \cmidrule(lr){6-8}
\multicolumn{2}{l}{} &
  \multicolumn{1}{c}{\textbf{Ball}} &
  \multicolumn{1}{c}{\textbf{Inner}} &
  \multicolumn{1}{c}{\textbf{Outer}} &
  \multicolumn{1}{c}{\textbf{Ball}} &
  \multicolumn{1}{c}{\textbf{Inner}} &
  \multicolumn{1}{c}{\textbf{Outer}} &
  \multicolumn{1}{c}{} \\[2pt] \midrule
\multicolumn{2}{l}{CNN9} & 
  81.2 \scriptsize$\pm$20.9 &
  75.1 \scriptsize$\pm$24.5 & 
  71.5 \scriptsize$\pm$26.9 &
  66.4 \scriptsize$\pm$20.3 & 
  81.3 \scriptsize$\pm$25.9 & 
  61.4 \scriptsize$\pm$40.1 & 
  72.8 \scriptsize$\pm$7.3 \\[2pt]
\multicolumn{2}{l}{CNN7} & 
  76.9 \scriptsize$\pm$21.3 & 
  65.9 \scriptsize$\pm$20.0 & 
  \textbf{88.9 \scriptsize$\pm$8.1} & 
  83.9 \scriptsize$\pm$15.7 & 
  85.3 \scriptsize$\pm$17.2 & 
  \textbf{93.5 \scriptsize$\pm$9.1} & 
  82.4 \scriptsize$\pm$8.9 \\[2pt]
\multicolumn{2}{l}{MobileNet V3 (large)} & 
  \textbf{96.7 \scriptsize$\pm$5.1} & 
  77.5 \scriptsize$\pm$21.8 & 
  76.5 \scriptsize$\pm$16.9 &
  88.9 \scriptsize$\pm$16.1 & 
  \textbf{97.6 \scriptsize$\pm$4.8} & 
  87.5 \scriptsize$\pm$20.0 &  
  \textbf{87.5 \scriptsize$\pm$8.3} \\[2pt]
\multicolumn{2}{l}{RegNet (x\_1\_6gf)} & 
  93.9 \scriptsize$\pm$7.1) & 
  \textbf{81.1 \scriptsize$\pm$19.7} & 
  74.1 \scriptsize$\pm$17.6 & 
  \textbf{92.0 \scriptsize$\pm$12.5} & 
  94.7 \scriptsize$\pm$9.3 & 
  84.0 \scriptsize$\pm$20.7 & 
  86.6 \scriptsize$\pm$7.6 \\[2pt] \bottomrule
\end{tabular}%
}
\end{table}

\begin{table}[ht]
\centering
\caption{100 $\times$ AUROC (mean $\pm$ std) for more complex/larger model architectures. Numbers in boldface indicate the best result for each column.}
\label{tab:larger-models}
{
\fontsize{9pt}{11pt}\selectfont
\tabcolsep5pt
\begin{tabular}{@{}lllllllll@{}}
\toprule
\multicolumn{2}{l}{\multirow{2.3}{*}{\textbf{Model}}} & 
\multicolumn{3}{c}{\textbf{Fan}} & 
\multicolumn{3}{c}{\textbf{Drive}} &
\multirow{2.3}{*}{\begin{tabular}[c]{@{}c@{}}\textbf{Macro} \\ \textbf{Average}\end{tabular}} \\ \cmidrule(lr){3-5} \cmidrule(lr){6-8}
\multicolumn{2}{l}{}  & 
  \textbf{Ball} & 
  \textbf{Inner} & 
  \textbf{Outer} & 
  \textbf{Ball} & 
  \textbf{Inner} & 
  \textbf{Outer} &\\[2pt] \midrule
\multicolumn{2}{l}{ConvNeXt (tiny)} & 
  \textbf{99.2 \scriptsize$\pm$3.2} & 
  65.5 \scriptsize$\pm$24.5 & 
  63.2 \scriptsize$\pm$31.0 &
  88.3 \scriptsize$\pm$16.7 & 
  94.2 \scriptsize$\pm$8.0 & 
  \textbf{88.4 \scriptsize$\pm$16.2} &  
  83.1 \scriptsize$\pm$13.8 \\[2pt]
\multicolumn{2}{l}{SwinTransformerV2 (tiny)} & 
  98.5 \scriptsize$\pm$3.9 & 
  77.7 \scriptsize$\pm$19.8 & 
  72.3 \scriptsize$\pm$22.5 & 
  86.2 \scriptsize$\pm$20.8 & 
  90.4 \scriptsize$\pm$12.6 & 
  80.2 \scriptsize$\pm$20.9 & 
  84.2 \scriptsize$\pm$8.6 \\[2pt]
\multicolumn{2}{l}{ViT (base)} & 
  91.8 \scriptsize$\pm$11.6 & 
  72.8 \scriptsize$\pm$23.4 & 
  \textbf{79.5 \scriptsize$\pm$24.6} & 
  86.6 \scriptsize$\pm$15.5 & 
  88.7 \scriptsize$\pm$14.3 & 
  87.2 \scriptsize$\pm$20.7 & 
  84.4 \scriptsize$\pm$6.4 \\[2pt]
\multicolumn{2}{l}{EfficientNetV2 (small)} & 
  94.6 \scriptsize$\pm$5.8 & 
  70.7 \scriptsize$\pm$22.3& 
  77.5 \scriptsize$\pm$22.2 & 
  \textbf{89.7 \scriptsize$\pm$16.1} & 
  97.2 \scriptsize$\pm$4.4 & 
  88.1 \scriptsize$\pm$21.0 & 
  \textbf{86.3 \scriptsize$\pm$9.3} \\[2pt]
\multicolumn{2}{l}{MaxViT} & 
  95.1 \scriptsize$\pm$6.4 & 
  \textbf{81.2 \scriptsize$\pm$17.1} & 
  70.2 \scriptsize$\pm$36.4 & 
  85.3 \scriptsize$\pm$17.9 & 
  \textbf{97.9 \scriptsize$\pm$3.1} & 
  87.1 \scriptsize$\pm$23.7 & 
  86.1 \scriptsize$\pm$9.1 \\[2pt] \bottomrule
\end{tabular}%
}
\end{table}

Another experiment was performed with architectures from the ResNet family with higher numbers of layers, namely ResNet34, ResNet50, and ResNet101. In Table \ref{tab:resnet-models}, it may be observed that although ResNet50 exhibited slightly better performance than that of ResNet18, this gain may be insufficient to justify its much longer training time due to a larger architecture. More importantly, the performance does not seem to steadily improve with increased complexity. Overall, these results suggest that, for the problem at hand, most gains are obtained not from increasing model complexity but rather from choosing a suitable architecture family, with the ResNet family appearing to be already the best choice among the candidates evaluated in this work.

\begin{table}[ht]
\centering
\caption{100 $\times$ AUROC (mean $\pm$ std) for other architectures within the RestNet family. Numbers in boldface indicate the best result for each column.}
\label{tab:resnet-models}
{
\fontsize{9pt}{11pt}\selectfont
\tabcolsep5pt
\begin{tabular}{@{}lllllllll@{}}
\toprule
\multicolumn{2}{l}{\multirow{2.3}{*}{\textbf{Model}}} &
  \multicolumn{3}{c}{\textbf{Fan}} &
  \multicolumn{3}{c}{\textbf{Drive}} &
  \multirow{2.3}{*}{\begin{tabular}[c]{@{}c@{}}\textbf{Macro} \\ \textbf{Average}\end{tabular}} \\ \cmidrule(lr){3-5} \cmidrule(lr){6-8}
\multicolumn{2}{l}{} &
  \multicolumn{1}{c}{\textbf{Ball}} &
  \multicolumn{1}{c}{\textbf{Inner}} &
  \multicolumn{1}{c}{\textbf{Outer}} &
  \multicolumn{1}{c}{\textbf{Ball}} &
  \multicolumn{1}{c}{\textbf{Inner}} &
  \multicolumn{1}{c}{\textbf{Outer}} &
  \multicolumn{1}{c}{} \\ \midrule
\multicolumn{2}{l}{ResNet34} &
  \textbf{98.9 \scriptsize$\pm$2.2} &
  82.0 \scriptsize$\pm$9.5 &
  86.0 \scriptsize$\pm$8.9 &
  85.8 \scriptsize$\pm$17.2 &
  94.4 \scriptsize$\pm$7.2 &
  89.6 \scriptsize$\pm$15.3 &
  89.4 \scriptsize$\pm$5.7 \\
\multicolumn{2}{l}{ResNet50} &
  97.1 \scriptsize$\pm$5.0) &
  \textbf{86.8 \scriptsize$\pm$11.7} &
  89.8 \scriptsize$\pm$11.2 &
  90.9 \scriptsize$\pm$11.3 &
  98.5 \scriptsize$\pm$4.3 &
  \textbf{91.6 \scriptsize$\pm$13.8} &
  \textbf{92.5 \scriptsize$\pm$4.1} \\
\multicolumn{2}{l}{ResNet101} &
  95.8 \scriptsize$\pm$7.2 &
  78.3 \scriptsize$\pm$15.0 &
  \textbf{90.1 \scriptsize$\pm$10.7} &
  \textbf{94.5 \scriptsize$\pm$9.6} &
  \textbf{98.7 \scriptsize$\pm$3.2} &
  90.2 \scriptsize$\pm$16.6 &
  91.2 \scriptsize$\pm$6.5 \\ \bottomrule
\end{tabular}%
}
\end{table}

Lastly, to address different representations other than the spectrogram, experiments with three different 1D signal representations were made with a smaller 1D CNN model architecture named WDCNN \cite{zhang2017new}. The experiments for the time, spectrum and power cepstrum signal representations can be seen in Table \ref{tab:sig-rep-exp}. The time and spectrum representations are commonly seen in the literature, and their results were not as good as those of the spectrogram on the ResNet model, with the spectrum being marginally better than the time representation. The power cepstrum is not as commonly seen as the other two representations, as cepstrum analysis is mostly used for gearbox fault diagnosis, but it can also be used for detecting the harmonics of bearing faults if they are well separated \cite{randall2021vibration}. Therefore, some other papers in the literature have attempted to use power cepstrum as input for deep learning models aiming to detect faults in the CWRU dataset \cite{bhakta2019fault, cai2021evaluation}. Table \ref{tab:sig-rep-exp} shows that the power cepstrum signal representation achieved a macro average AUROC of 0.897, which, although inferior to that of the spectrogram on the ResNet model, is still quite competitive, as the WDCNN is much smaller, trains faster, and needs a smaller data segment than the ResNet models.


\begin{table}[ht]
\centering
\caption{
100 $\times$ AUROC (mean $\pm$ std) for experiments using the WDCNN and ResNet18 architecture on different signal representations. Numbers in boldface indicate the best result for each column.}
\label{tab:sig-rep-exp}
{
\fontsize{9pt}{11pt}\selectfont

\tabcolsep4pt
\begin{tabular}{@{}lllllllll@{}}
\toprule
\multirow{2.3}{*}{\textbf{Model}} &
  \multirow{2.3}{*}{\begin{tabular}[c]{@{}l@{}}\textbf{Signal} \\ \textbf{Representation}\end{tabular}} &
  \multicolumn{3}{c}{\textbf{Fan}} &
  \multicolumn{3}{c}{\textbf{Drive}} &
  \multirow{2.3}{*}{\begin{tabular}[c]{@{}c@{}}\textbf{Macro} \\ \textbf{Average}\end{tabular}} \\ \cmidrule(lr){3-5} \cmidrule(lr){6-8}
 &
   &
  \multicolumn{1}{c}{\textbf{Ball}} &
  \multicolumn{1}{c}{\textbf{Inner}} &
  \multicolumn{1}{c}{\textbf{Outer}} &
  \multicolumn{1}{c}{\textbf{Ball}} &
  \multicolumn{1}{c}{\textbf{Inner}} &
  \multicolumn{1}{c}{\textbf{Outer}} &
   \\
   \midrule
\multirow{3.8}{*}{WDCNN} &
  Time &
  91.9 \scriptsize$\pm$7.7 &
  68.4 \scriptsize$\pm$18.8 &
  82.7 \scriptsize$\pm$7.1 &
  68.6 \scriptsize$\pm$25.6 &
  83.7 \scriptsize$\pm$21.9 &
  81.0 \scriptsize$\pm$18.8 &
  79.4 \scriptsize$\pm$8.4 \\[4pt]
&
  Spectrum &
  85.0 \scriptsize$\pm$12.2 &
  68.3 \scriptsize$\pm$16.7 &
  83.0 \scriptsize$\pm$7.1 &
  85.5 \scriptsize$\pm$18.0 &
  85.5 \scriptsize$\pm$17.3 &
  \textbf{88.9 \scriptsize$\pm$10.0} &
  82.7 \scriptsize$\pm$6.7 \\[4pt]
 &
   Power Cepstrum & 
   93.8 \scriptsize$\pm$6.3 &
   \textbf{82.3 \scriptsize$\pm$10.8} &
   \textbf{89.5 \scriptsize$\pm$8.7} &
   97.5 \scriptsize$\pm$2.3 &
   \textbf{95.4 \scriptsize$\pm$4.9} &
   79.4 \scriptsize$\pm$20.0 &
   \textbf{89.7 \scriptsize$\pm$6.7}
   \\
  \bottomrule
\end{tabular}%
}
\end{table}

\subsection{Fault detection}
\label{subsec:fault-detection}

As mentioned in Section~\ref{subsubsec:error-analysis}, while some confusion between fault types occurs, it is mostly restricted to fault types within the same location. This suggests that the model may be effectively used for the simpler problem of detecting when a fault is present at a specific location, i.e., the model should output a positive classification if and only if a fault (of any type) is present at the same location where the signal fed to model is acquired. This may be seen as the conventional problem of fault detection (without diagnosis), except that localizing the fault is implicitly performed, i.e., whenever a fault is detected by this method, its location is already identified.

To evaluate this fault detection problem,
the three fault type labels were condensed (by the max operation) into a single binary label indicating whether a fault has occurred at that location. Correspondingly, the model was modified by taking the 
average\footnote{We also experimented with taking the maximum of the probability output, but the results were not improved.} of the probability output of the three fault-type detectors (akin to a soft-voting ensemble). 
In other words, for a given sample, at a specific location, the predicted probability of a fault is computed as $\hat{p}_F = (\hat{p}_I + \hat{p}_O + \hat{p}_B)/3$, where $\hat{p}_I$, $\hat{p}_O$ and $\hat{p}_B$ denote, respectively, the corresponding predicted probabilities of inner, outer and ball faults.

\begin{figure*}[t]
    \centering
    \includegraphics[scale=0.5]{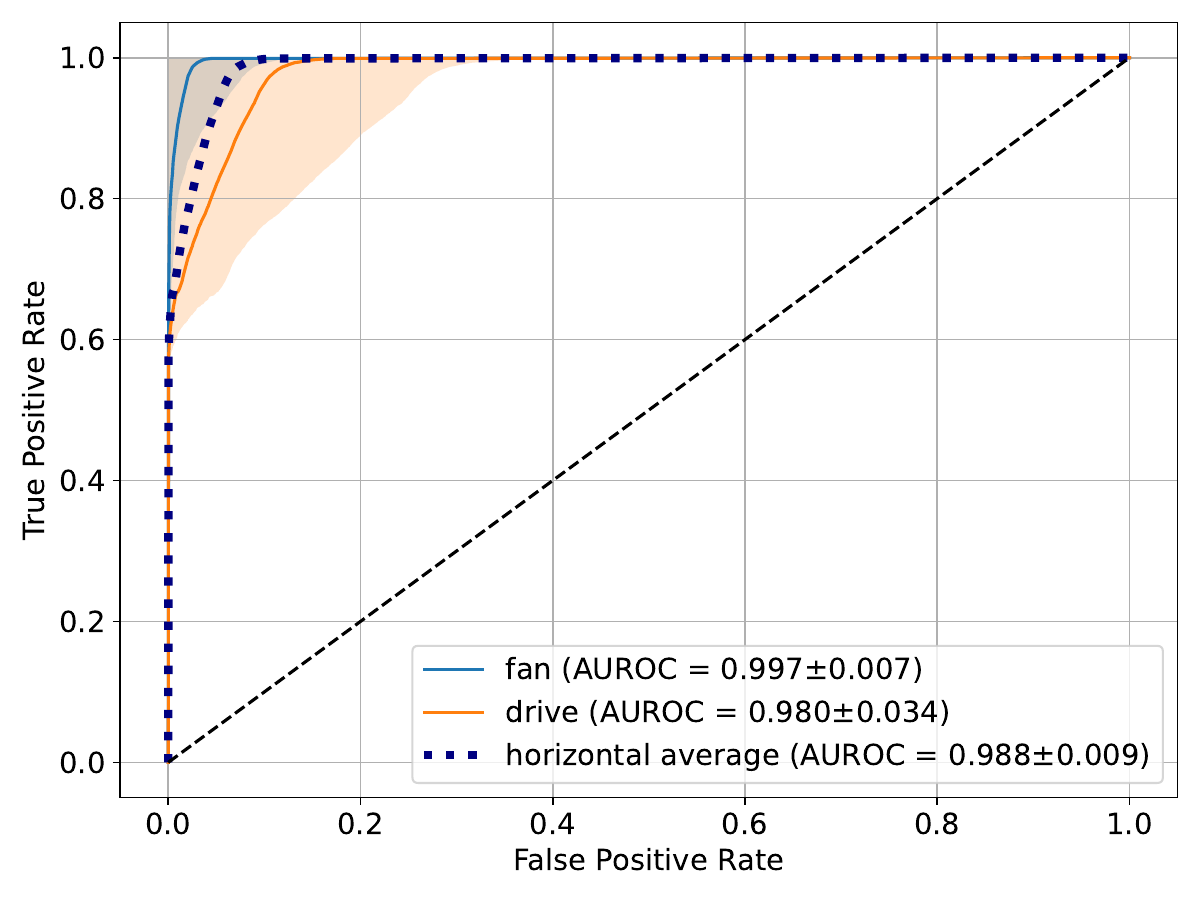}
    \caption{ROC curve for fault detection at each location. Solid curves represent the average ROC curve across 30 realizations, while the filled region represents the standard deviation.} 
    \label{fig:location_model}
\end{figure*}

Figure \ref{fig:location_model} confirms the observation made from the logits boxplots that the model does indeed perform very well at detecting faults at their locations, with a macro average AUROC of $0.988\pm0.009$.
(Note that the lower performance of the drive-end detector can be explained the peculiarities of the drive-inner-14 and drive-outer-14 conditions, as discussed in Section~\ref{subsubsec:error-analysis}.)
Considering the real-world use of these models, where the crucial, most important information is whether a component
is faulty and needs to be replaced, these results seem promising. 
While the fault-type information is still important for interpretability and trust in the model predictions, as well as for the decision on when to change the bearings, reliably detecting when and where a bearing fault occurs (without finer diagnosis) may already be useful in practice, where numerous fault types other than bearing faults happen and differentiating what is a bearing fault aggregates significant value.

It should be emphasized that the fault detection problem cannot be formulated without leakage on the CWRU dataset using a multi-class approach since the healthy class (containing a single-bearing configuration) will necessarily have to be split between train and test sets. Only a multi-label formulation that places all signals from the healthy state on the test set, as pioneered by this paper, enables a rigorous evaluation of machine learning models for fault detection using the CWRU dataset.

\section{Discussion and conclusion}
\label{sec:conclusions}

In this paper, we propose several modifications to the standard formulation of the bearing fault detection and diagnosis problem using the CWRU dataset in order to bring it closer to real-life industry settings. Hendriks et al. \cite{hendriks2022towards} (and, more recently, also Abburi et al. \cite{abburi2023closer}) have demonstrated that bearing data leakage profoundly affects the performance of machine learning models, posing a serious shortcoming when the goal is to design algorithms that can detect and diagnose faults in different bearings than those used for training. However, their work has not resonated enough within the published literature, and most papers still use a dataset division containing data leakage, generating over-optimistic results. 

Our paper takes their approach a step further by proposing a multi-label problem formulation that allows for the detection of different fault types occurring simultaneously at each location, without the need for synchronous acquired signals, while keeping it completely data-leakage-free. By construction, the proposed methodology completely solves the healthy bearing data leakage and imbalance and ensures that the model evaluation is solid and reliable while more accurately reflecting real-world conditions. 

Consolidating our framework, we explored various training approaches to enhance model performance. This resulted in our proposed dataset division that takes greater advantage of the CWRU dataset by expanding the train/test split ratio from 1:2 to 2:1 and adding diversity with random fault size configurations for each fault location and type pair. Additionally, we proposed dividing the fault type and location detection into two different problems while using only a single model, effectively doubling our training dataset. In our ablation studies, these proposed approaches have demonstrated significant performance improvements, with the macro average AUROC going from 0.781 to 0.911.

As an application of our approach, we conducted a comparative benchmark using several model architectures of different sizes and levels of complexity, as well as different signal representations as the input to the network. None of the architectures/representations evaluated achieved significant gains over the ResNet18 applied to spectrogram images, although the WDCNN applied to the power cepstrum obtained competitive results considering its lower computational requirements.

It is worth mentioning that, among previous works applying deep neural networks to the CWRU dataset, we have found many that do not describe their hyperparameter optimization procedure, raising the possibility that it was done using test set performance as the objective, which would result in a biased evaluation. Thus, another contribution of our work comes from the specification and use of a model selection and evaluation procedure intended to minimize such bias.

After fine-tuning the model, we conducted an error analysis that generated interesting insights. We noticed that our model struggled to differentiate between inner and outer faults and had difficulty detecting certain specific fault conditions. However, these errors could mostly be explained by certain aspects of the CWRU dataset discussed in Smith and Randall's thorough analysis of the dataset \cite{smith2015rolling}. This adds to the credibility of our results, as one should always be skeptical of surprisingly good results that diverge significantly from what would be expected after a careful analysis by domain experts.

In conclusion, the CWRU  dataset remains an important resource for research on rolling bearing fault detection and diagnosis and will likely continue to be widely utilized. We recommend that future researchers using the CWRU dataset with machine learning models for fault detection and diagnosis use our proposed framework, as it offers a realistic and reliable evaluation. 



\section*{\textit{CRediT authorship contribution statement}}

\textbf{Rodrigo Kobashikawa Rosa:} Conceptualization, Data curation, Formal analysis, Visualization, Software, Writing - original draft. \textbf{Danilo Braga:} Resources, Supervision, Writing - review \& editing. \textbf{Danilo Silva:} Conceptualization, Methodology, Project administration, Supervision, Writing - review \& editing. 

\section*{Acknowledgements}

This research was supported by the Conselho Nacional de Desenvolvimento Científico e Tecnológico (CNPq) through grants 164299/2021-1 and 304619/2022-1.

 \bibliographystyle{elsarticle-num} 
 \bibliography{refs}





\end{document}